# Surface Tension Supported Floating of Heavy Objects: Why Elongated Bodies Float Better?

Edward Bormashenko

*Ariel University, Physics Department, Ariel, POB 3, 40700, Israel*

E-mail address: edward@ariel.ac.il

**Abstract**

Floating of bodies heavier than the supporting liquid is discussed. Floating of cylindrical, ellipsoidal bodies and rectangular plates possessing lateral dimensions smaller than the capillary length is treated. It is demonstrated that more elongated bodies of a fixed volume are better supported by capillary forces, due to the increase in the perimeter of the triple line. Thus, floating of metallic needles obtains reasonable explanation.

**1. Introduction**

Physics of floating as many other fundamental physical phenomena was first studied by Archimedes. For the details of the detective story of the revealing and restoring the original text of the Archimedes Palimpsest containing the treatise devoted to the floating of bodies see Refs. 1–2.

The famous Archimedes principle formulated in the Palimpsest states that: "any body wholly or partially immersed in a fluid experiences an upward force (buoyancy) equal to, but opposite to the weight of the fluid displaced". It seems that it follows from this principle that an object can float only if is less dense than the liquid in which it is placed. However, a sure-handed child may place a steel needle on a surface of water, and it will float, as shown in **Fig. 1**. The first explanation of this effect belongs to Galileo Galilei [3–4]. Consider heavy plate placed on a water surface, as shown in **Fig. 2**. When the three-phase (triple line) is firmly pinned to the surface of the plate, it may displace a volume which is much larger than the total volume of the plate itself, as shown in **Fig. 3**. Hence the buoyancy will be essentially increased (it is noteworthy that the effect of pinning of the triple line is responsible on a broad diversity of wetting phenomena ([5–9]). It is seen that Galilei related floating of heavy bodies to the increase of the Archimedes force only, and this explanation is at least partially true.

The unbelievable physical intuition of Galilei is admirable, but actually floating of heavy objects arises as the interplay of the buoyancy and surface tension. The restoring force that counteracts the floating body weight *mg*, comprises a surface tension force and the buoyancy force arising from hydrostatic pressure, as shown in **Fig. 3A**. Following the Galilei idea, illustrated with **Fig. 3A** the buoyancy equals to the total weight of the liquid displaced by the body, shown with the dark-gray in **Fig. 3A**. Hand in hand with the buoyancy the surface tension supports floating as shown in **Fig. 3A-B**. Keller demonstrated [10] that the total force supporting the floating equals the weight of the entire volume of liquid displaced by the body, as depicted in **Fig. 3B**. The important problem of the maximal depth of sinking of a floating object is discussed in detail in Ref. 4.

The interest to the floating of bodies heavier than the supporting liquid was revived recently [11–14] due to a variety of effects, including the ability of water striders to walk on water [15–17] and self-assembly of floating particles [18–19]. An understanding of floating is crucial for a variety of biological and engineering problems, including the behavior of colloidal particles attached to a liquid surface, the formation of lipid droplets and liquid lenses, etc. [20–23].

Everybody who tried to place an object heavier than water on its surface knows that it is more convenient to do this with prolonged objects. It is much simpler to do this procedure with a steel needle than with a steel sphere of the same mass. Thus, we address the question: why elongated bodies float better? We restrict our treatment with small bodies possessing a characteristic lateral dimension smaller than the so-called capillary length, describing the interrelation between buoyancy and capillarity and is defined by the expression: $l_{ca} = \sqrt{\gamma / \rho_L g}$ (where $\gamma$ and $\rho_L$ are the surface tension and density of the liquid respectively) [24–25]. When the characteristic lateral dimension of floating bodies is much less than the capillary length, the buoyancy is negligible, and floating is totally prescribed by the surface tension [4]. It is noteworthy that the capillary length is on the same order of magnitude of several millimeters for all liquids [24]. The paper proposes a simple qualitative model clarifying the impact of the shape on the floating of heavy bodies.

## 2. Discussion

### 2.1. Floating of heavy cylindrical needles

Consider floating of the cylindrical body which radius is smaller than the capillary length. Thus the effects due to the buoyancy may be neglected. The gravity effects $f_{grav}$ are obviously given by:

$$f_{grav} = \rho g V = \frac{\pi}{4} g \rho d^2 l, \qquad (1)$$

where $\rho$, $V$, $l$ and $d$ are the density, volume, length and diameter of the needle respectively (see **Fig. 4**). The capillary force withstanding gravity is estimated as: $f_{cap} \cong \gamma \xi f(\theta)$ where $\xi$ is the perimeter of the triple (three-phase) line, and $f(\theta)$ is the function depending on the apparent contact angle $\theta$ [4, 14]. The *maximal* capillary force $f_{cap}^{max}$ withstanding gravity and supporting floating may be very roughly estimated as:

$$f_{cap}^{max} \cong \gamma \xi_{max} = 2\gamma(d + l), \qquad (2)$$

where $\xi_{max}$ is the maximal perimeter of the triple line, corresponding to the cross-section ABCD (see **Fig. 4**). This capillary force is maximal when it has only a vertical component (this is possible for hydrophobic surfaces, when the apparent contact angle characterizing the triad solid/liquid/vapor is larger than $\frac{\pi}{2}$, as shown in Ref. 4), and a liquid wets a cylindrical needle along a line dividing the floating body into equal parts (ABCD is the medial longitudinal cross-section of the needle, bounded by the triple line), as shown in **Fig. 4**. The interrelation between the maximal capillarity and gravity-induced effects will be described by the dimensionless number:

$$\Psi = \frac{f_{cap}^{max}}{f_{grav}}. \qquad (3)$$

Substituting expressions (1) and (2) into (3) yields:

$$\Psi = \frac{f_{cap}^{max}}{f_{grav}} \cong \alpha \frac{d + l}{d^2 l}, \qquad (4)$$

where

$$\alpha = \frac{8\gamma}{\pi g \rho}.$$

It should be mentioned that the dimensionless number $\Psi$ is actually the inverse of the well-known Eötvös (or Bond) number [13, 14]. Now consider floating of various cylindrical needles of the same volume. Thus, Eq. (5) takes place:

$$\frac{\pi}{4}d^2 l = V = \text{const}. \tag{5}$$

Hence, the length of the needle may be expressed as $l = 4V/\pi d^2$. Substituting this expression into formula (4) and considering the constancy of volume (5) yields for $\Psi$:

$$\Psi(d) \cong \frac{\pi\alpha}{V}\left(d + \frac{4V}{\pi d^2}\right). \tag{6}$$

The function $\Psi(d)$ is schematically depicted in **Fig. 5**. It possesses a minimum, when $d = d^* = 2\sqrt[3]{V/\pi}$. Considering $V = (\pi/4)d^2 l$ gives rise to $d^* = 2l$. It means that the influence of capillary forces supporting floating is minimal for needles possessing close longitudinal and lateral dimensions, whereas this influence is maximal for very long ($d^* \ll 2l$) and very short ($d^* \gg l$) needles. Somewhat curiously, very short needles could also be treated as oblong (elongated) objects, and the inequality $\Psi \gg 1$ takes place. We conclude that the relative influence of capillary forces is maximal for elongated objects (in turn, the relative influence of gravity for these objects is minimal). This consideration qualitatively explains the floating ability of heavy needles.

### 2.2. Floating of rectangular plates

Floating of heavy ($\rho > \rho_L$) rectangular plates may be treated in a similar way. Consider floating a rectangular plate with the thickness of $h$ and lateral dimensions $a$x$b$. The gravity in this case is given by:

$$f_{\text{grav}} \cong \rho a b h g . \tag{7}$$

The maximal capillary force is estimated as:

$$f_{\text{cap}}^{\max} \cong \gamma\xi = 2\gamma(a+b), \tag{8}$$

where $\xi$ is the perimeter of the cross-section ABCD (see **Fig. 6**). The capillary force is maximal when it has only a vertical component (the cross sections for rectangular plates are the same). Consider rectangular plates of the same thickness $h$ and the same volume $V=abh$, but differently shaped. We have for the dimensionless number $\Psi$:

$$\Psi = \frac{f_{cap}^{max}}{f_{grav}} \cong \frac{2\gamma(a+b)}{\rho g V} = \alpha'(a+b), \qquad (9)$$

where

$$\alpha' = \frac{2\gamma}{\rho g V}.$$

Expression (9) could be rewritten as:

$$\Psi(a) = \alpha'\left(a + \frac{V}{ah}\right). \qquad (10)$$

Function $\Psi(a)$ is qualitatively similar to that depicted in **Fig. 5**. It possesses a minimum, when $a = a^* = \sqrt{V/h}$.

Considering $V = abh$ yields $a^* = b$. Thus, the influence of capillary forces is minimal for square plates of a fixed thickness and fixed volume. Again, the influence of capillarity supporting the floating is maximal for elongated rectangular plates.

### 2.3. Floating of ellipsoidal objects

Treatment of ellipsoidal objects is more complicated. Consider a heavy completely hydrophobic ($\theta=180°$) ellipsoidal body possessing the semi-principal axes of length $a$, $b$, $c$, floating as shown in **Fig. 7** (in the shown case the capillary force is maximal; it is also supposed that it is vertical). The gravitational force acting on ellipsoid is given by:

$$f_{grav} \cong \frac{4}{3}\pi\rho abcg, \qquad (11)$$

where $(4/3)\pi abc = V$ is the volume of the ellipsoid. The maximal capillary force is estimated as $f_{cap}^{max} \cong \gamma\xi_{max}$, where $\xi_{max}$ is the maximal perimeter of the triple line, corresponding to the circumference of the ellipse, depicted in red in **Fig. 7**. The circumference of the ellipse is the complete elliptic integral of the second kind. It may be reasonably evaluated with the approximate Ramanujan formula: $\xi_{max} \cong \pi[3(a+b) - \sqrt{10ab + 3(a^2 + b^2)}]$. Thus, the dimensionless number $\Psi$ equals:

$$\Psi = \frac{f_{cap}^{max}}{f_{grav}} \cong \alpha''[3(a+b) - \sqrt{10ab + 3(a^2 + b^2)}], \qquad (12)$$

where

$$\alpha'' = \frac{\gamma\pi}{\rho g V}.$$

Now we fix the values of the volume $V$ and the vertical semi-axis $c$. Thus, we obtain:

$$b = \frac{3V}{4\pi ac} = \frac{\beta}{a}; \beta = \frac{3V}{4\pi c}. \quad (13)$$

Substituting expression (13) into (12) yields:

$$\Psi(a) \cong \alpha'' \left[ 3\left(a + \frac{\beta}{a}\right) - \sqrt{10\beta + 3\left(a^2 + \frac{\beta^2}{a^2}\right)} \right]. \quad (14)$$

Differentiation of equation (14) yields the somewhat cumbersome expression:

$$\frac{d\Psi(a)}{da} \cong 3\alpha'' \left(1 - \frac{\beta}{a^2}\right) \left( 1 - \frac{a\left(1 + \frac{\beta}{a^2}\right)}{\sqrt{10\beta + 3\left(a^2 + \frac{\beta^2}{a^2}\right)}} \right). \quad (15)$$

However, it could be easily demonstrated that the expression

$$1 - \frac{a\left(1 + \frac{\beta}{a^2}\right)}{\sqrt{10\beta + 3\left(a^2 + \frac{\beta^2}{a^2}\right)}}$$

is always positive; hence $\frac{d\Psi(a)}{da} = 0$ takes place when $a = \pm\sqrt{\beta}$. Hence, the function $\Psi(a)$ has a physically meaningful minimum, when $a = \sqrt{\beta} = \sqrt{3V/4\pi c} = \sqrt{ab}$. Finally, this results in $a = b$. We conclude that the role of capillary forces is minimal in the degenerated case, when horizontal semi-axes of an ellipsoid are the same, namely the cross-section bounded by the triple line is a circle. This result is expected intuitively from the variational considerations. Indeed, we fixed the volume $V$ and the vertical semi-axis $c$ of the floating ellipsoid. In this situation, the area of the cross-section bounded by the triple line $S = \pi ab$ turns out to be fixed (recall, that $V = (4/3)\pi abc = (4/3)Sc = \text{const}$). Thus, we are actually seeking the minimal circumference of the figure possessing the given area, thereby supplying the minimum to the capillary force. It is well-known from the variational analysis that a circle has the minimum possible perimeter for a given area. Curiously, the use of the approximated Ramanujan formula for the calculation of the ellipse perimeter leads to an accurate solution. We come to the conclusion that the more elongated ellipsoidal body is better supported by capillary forces. The approach presented in the paper is qualitative; we did not calculate carefully neither gravity nor surface tension induced effects, as it was performed in references [4, 14, 26]. However this qualitative approach, based on very crude approximations, explains the fascinating floating ability of heavy elongated objects.

## 3. Conclusions

Floating of heavy bodies possessing characteristic lateral dimensions smaller than the capillary length is discussed. In this case the buoyancy is negligible, and floating is totally prescribed by the surface tension related effects. The maximal possible capillary force, supporting floating is proportional to the largest perimeter of the cross-section of the body circumscribed by the triple line. This perimeter increases with the elongation for a given volume of a body. This conclusion is true for different symmetries of floating objects, including cylindrical and ellipsoidal bodies and rectangular plates (for rectangular plates and ellipsoidal objects vertical dimensions are also fixed). These arguments qualitatively explain the floating of metallic needles (whereas a metallic ball of the same mass will sink).

**Acknowledgements**

I am thankful to Mrs. Ye. Bormashenko and Mrs. A. Musin for their kind help in preparing this manuscript. I am grateful to Professor G. Whyman for inspiring discussions. I am thankful to Dr. R. Grynyov for experiments demonstrating floating of heavy objects.

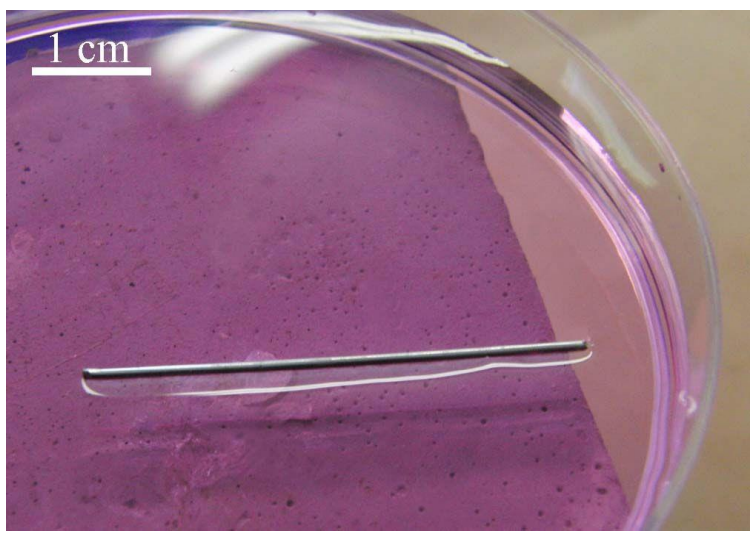

**Fig. 1.** Steel needle floating on the water surface.

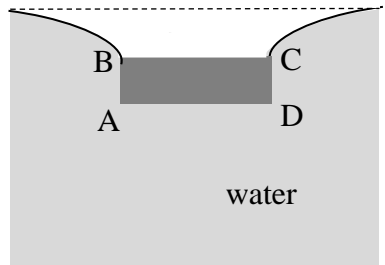

**Fig. 2**. Galileo intuitive reasoning explaining the floating of the heavy plate ABCD. When the triple line is pinned to the surface of the plate, it may displace the volume which is much larger than the total volume of the plate itself.

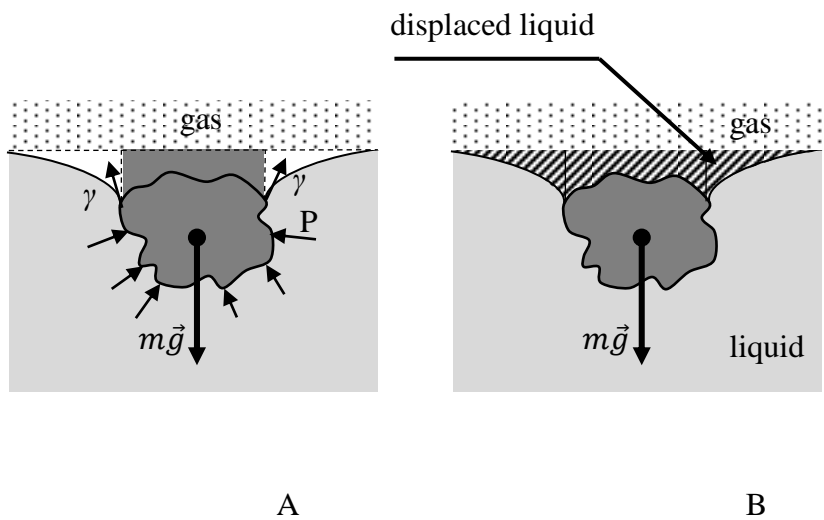

A  B

**Fig. 3**. **A**. The buoyancy of the floating heavy object equals to the total weight of the displaced water, shown shaded, dark-gray. The surface tension force which equals $\gamma$ for the unit length of the triple line also supports floating. **B**. The total restoring force including the surface tension and buoyancy equals the total weight of the displaced liquid (the gray shaded area in **Fig. 3.2 B**).

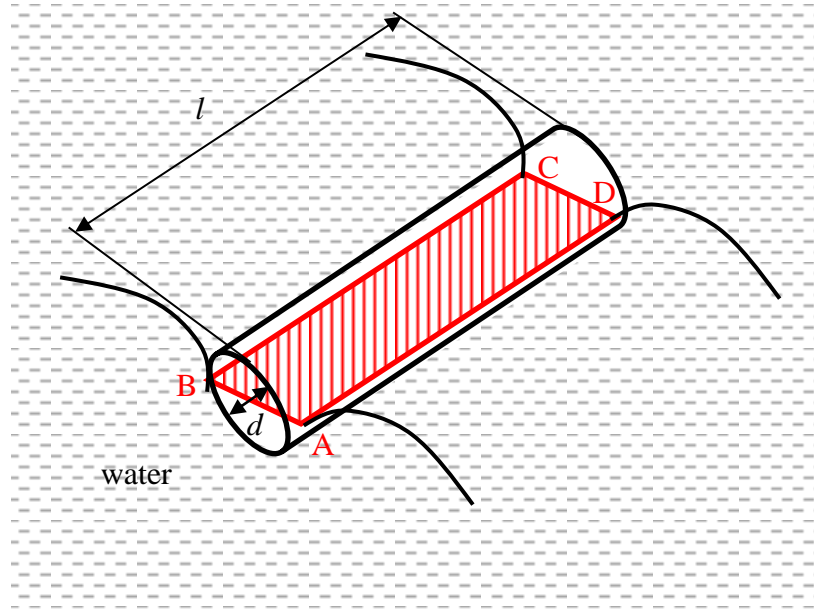

**Fig. 4.** Scheme of a floating cylindrical needle supported by the surface tension. ABCD is the cross-section bounded by the triple (three-phase) line.

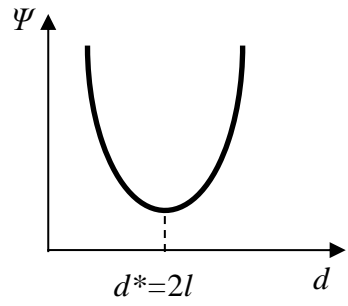

**Fig. 5**. The dimensionless number $\Psi = \dfrac{f_{cap}^{max}}{f_{grav}}$ depicted schematically as a function of the diameter of the needle.

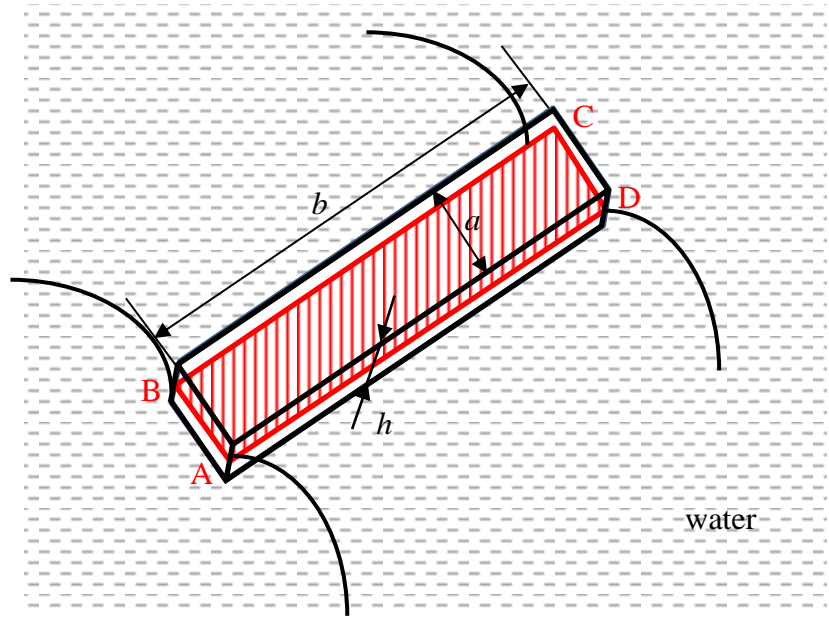

**Fig. 6.** Floating rectangular plate with the dimensions of *a*x*b*x*h*.

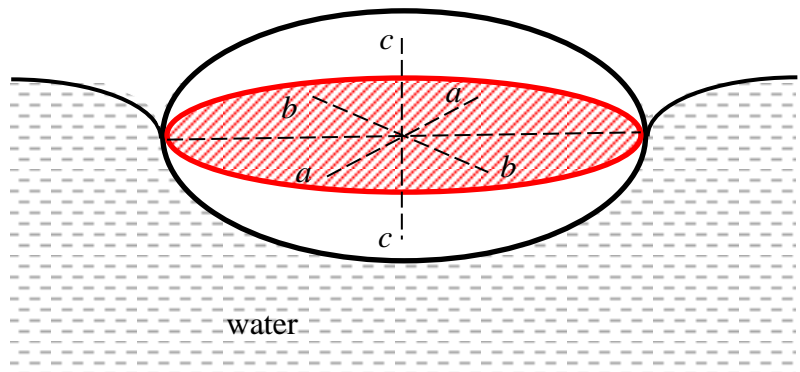

**Fig. 7.** Floating ellipsoidal object; *a*,*b*, *c* are semi-axes. The cross-section circumscribed by the triple line is shown in red.

**Figures legends**

**Fig. 1.** Steel needle floating on the water surface.

**Fig. 2**. Galileo intuitive reasoning explaining the floating of the heavy plate ABCD. When the triple line is pinned to the surface of the plate, it may displace the volume which is much larger than the total volume of the plate itself.

**Fig. 3**. A. The buoyancy of the floating heavy object equals to the total weight of the displaced water, shown shaded, dark-gray. The surface tension force which equals $\gamma$ for the unit length of the triple line also supports floating. B. The total restoring force including the surface tension and buoyancy equals the total weight of the displaced liquid (the gray shaded area in Figure 3.2 B).

**Fig. 4**. Scheme of a floating needle supported by the surface tension. ABCD is the cross-section bounded by the triple (three-phase) line.

**Fig. 5**. The dimensionless number $\Psi = \dfrac{f_{cap}^{max}}{f_{grav}}$ depicted schematically as a function of the diameter of the needle.

**Fig. 6.** Floating rectangular plate with the dimensions of $a \times b \times h$.

**Fig. 7.** Floating ellipsoidal object; $a$, $b$, $c$ are semi-axes. The cross-section circumscribed by the triple line is shown in red.

**Highlights**

*Floating of bodies heavier than the supporting liquid is discussed.

*Floating of objects possessing lateral dimensions smaller than the capillary length is treated.

*Floating these bodies is prescribed by the surface tension related effects.

*Elongated bodies of a fixed volume are better supported by capillary forces, due to the increase in the perimeter of the triple line.